\documentclass[12pt,showpacs,showkeys,amsmath,amssymb]{revtex4}
\usepackage{amsmath,amsfonts,amsthm,amscd,amssymb,latexsym}
\usepackage{bm}
\usepackage{dcolumn}
\usepackage{graphicx}
\usepackage{epstopdf}
\usepackage{color}
\usepackage{epsf}
\usepackage{epsfig}
\usepackage{graphicx, epic, eepic, color}

\newcommand{\beq}{\begin{equation}}
\newcommand{\eeq}{\end{equation}}

\begin{document}

\title{Toward Nonlocal Electrodynamics of Accelerated Systems}

\author{Bahram \surname{Mashhoon}$^{1,2}$}
\email{mashhoonb@missouri.edu}

\affiliation{$^1$Department of Physics and Astronomy, University of Missouri, Columbia, Missouri 65211, USA\\
$^2$School of Astronomy, Institute for Research in Fundamental
Sciences (IPM), P. O. Box 19395-5531, Tehran, Iran\\
}

\date{\today}

\begin{abstract}
We revisit acceleration-induced nonlocal electrodynamics and the phenomenon of photon spin-rotation coupling. The kernel of the theory for the electromagnetic field tensor involves parity violation under the assumption of linearity of the field kernel in the acceleration tensor. However, we show that parity conservation can be maintained by extending the field kernel to include quadratic terms in the acceleration tensor. The field kernel must vanish in the absence of acceleration; otherwise, a general dependence of the kernel on the acceleration tensor cannot be theoretically excluded.  The physical implications of the quadratic kernel are briefly discussed.
\end{abstract}

\pacs{03.30+p, 11.10.Lm, 04.20.Cv}
\keywords{relativity, accelerated systems, nonlocal electrodynamics}

\maketitle

\section{Introduction}

Consider an inertial observer $\mathcal{O}$ moving with constant velocity $\mathbf{v}$ in an inertial frame of reference in Minkowski spacetime. 
The observer's 4-velocity $u^\mu = dx^\mu/d\tau$
can be expressed in Cartesian coordinates $x^\mu = (t, x, y, z)$ as $u^\mu = \gamma (1, \mathbf{v})$, where $\tau$ is the observer's proper time, $d\tau = dt/\gamma$, and $\gamma$ is the Lorentz factor. The observer measures the frequency $\omega$ of an incident plane monochromatic electromagnetic wave with propagation vector $k_0^{\mu} = (\omega_0, \mathbf{k}_0)$, where $\omega_0$ is the frequency and $\mathbf{k}_0$ is the wave vector with $\omega_0 =  |\mathbf{k}_0|$, as measured by inertial observers $\mathcal{O}_0$ at rest in the background inertial frame. As is well known, $\omega$ is given by the relativistic Doppler formula, namely, 
\begin{equation}\label{I1}
\omega = - \eta_{\mu \nu}\,k_0^\mu\,u^\nu = \gamma \,(\omega_0 - \mathbf{v} \cdot \mathbf{k}_0)\,.
\end{equation}
Throughout this paper, we use the convention that c = 1, unless specified otherwise; moreover, Greek indices run from 0 to 3, while Latin indices run from 1 to 3. The Minkowski metric tensor $\eta_{\mu \nu}$ is given by diag$(-1, 1, 1, 1)$. 

Let us note that Eq.~\eqref{I1} can be expressed as $-\omega_0\,dt +\mathbf{k}_0 \cdot d\mathbf{x} = -\omega\, d\tau$, which follows from the invariance of the phase of the wave under Lorentz transformation. Indeed, if we consider the class of moving inertial observers all at rest in the inertial frame of $\mathcal{O}$, we recover from the Lorentz invariance of the phase the relativistic relations for both the frequency $\omega$ and the corresponding wave vector $\mathbf{k}$, namely,  
\begin{equation}\label{I2}
\mathbf{k} = \mathbf{k}_0 +\frac{1}{v^2} (\gamma - 1)(\mathbf{v} \cdot \mathbf{k}_0)\,\mathbf{v} - \gamma\,\mathbf{v}\, \omega_0\,,
\end{equation}
where $v = |\mathbf{v}|$. 

The Lorentz invariance of the phase is a consequence of a more basic approach that deals with the electromagnetic radiation field in the two Lorentz frames under consideration here. Transforming the radiation field, either the gauge field $A_\mu$ or the field tensor $F_{\mu \nu} = \partial_\mu A_\nu - \partial_\nu A_\mu$, from the background inertial frame to the moving inertial frame results in the invariance of the phase of the radiation under Lorentz transformation.  

We now wish to extend our treatment to accelerated observers, since actual observers are all more or less accelerated. Let us for the moment extend the theory of relativity to such observers in the standard manner by applying Lorentz transformations point by point along their accelerated world lines~\cite{AE}. The pointwise extension of Lorentz invariance is the content of the hypothesis of locality that is at the foundation of the special and general theories of relativity. The locality assumption implies that an accelerated observer is, at each event along its world line, locally equivalent to an otherwise identical momentarily comoving inertial observer. To avoid unphysical situations when dealing with accelerated motion, we generally assume that the acceleration is turned on at some initial time $t_i$ and, after a finite interval of time, the acceleration is turned off at $t_f$; moreover, for $t<t_i$ and $t>t_f$, the observer moves uniformly in an inertial frame of reference. In most laboratory experiments, the accelerations are small and the locality postulate is a good approximation; however, in situations involving high accelerations the locality assumption may not be adequate~\cite{BM1, BM2, Maluf:2010fb, Maluf:2011gt, BMB}.

By way of illustration, let us consider a thought experiment involving an observer that for $t \ge 0$ rotates uniformly about the $z$ axis in the $(x, y)$ plane of an inertial frame of reference. More specifically, for $-\infty <t<0$, the observer under consideration moves uniformly along a straight line parallel to the $y$ axis with speed $r\,\Omega_0$ in the $(x, y)$ plane; that is, $x = r > 0$ and $y=r\,\Omega_0\,t$.  At $t=0$, the observer is forced to move counterclockwise with uniform angular velocity $\Omega_0 >0$ on a circular orbit of radius $r$ about the $z$ axis. Thus, for $t\ge 0$, $x=r\,\cos \varphi$ and $y=r\,\sin \varphi$, where $\varphi =\Omega_0\,t$. The rotating observer with proper time $\tau = t /\gamma$ measures the frequency  of a plane monochromatic electromagnetic wave that propagates along the $z$ axis. In this case of normal incidence, the locality postulate leads to the local invariance of the phase and we find from Eq.~\eqref{I1},
\begin{equation}\label{I3}
\omega_{\rm D} = \gamma\,\omega_0\,, \qquad \gamma = (1-v^2)^{-1/2}\,, \qquad v = r\,\Omega_0\,, 
\end{equation}
which is the expression for the transverse Doppler effect. This result can be physically interpreted in terms of the phenomenon of time dilation. On the other hand, transforming the radiation field to the rotating frame results in Fourier components with frequencies 
\begin{equation}\label{I4}
{\tilde \omega}_{\pm} = \gamma\,(\omega_0 \mp \Omega_0)\,.
\end{equation}
The actual calculation involved in this more general result can be carried out using the explicit transformation of either $A_\mu$ or $F_{\mu \nu}$ to the rotating frame~\cite{BM3} or by means of projecting these fields on the natural orthonormal tetrad frame  of the rotating observer~\cite{HaMa}. An observer generally carries an adapted orthonormal tetrad frame field $\lambda^{\mu}{}_{\hat \alpha}$ such that 
\begin{equation}\label{I4a}
\eta_{\mu \nu}\,\lambda^{\mu}{}_{\hat \alpha}\,\lambda^{\nu}{}_{\hat \beta} =  \eta_{\hat \alpha \hat \beta}\,, \qquad \lambda^{\mu}{}_{\hat 0} = \frac{dx^\mu}{d\tau}\,.
\end{equation}
For $t\ge 0$, the components of the rotating observer's tetrad frame in 
$(t, x, y, z)$ coordinates are given by
\begin{align}
\label{I5}\lambda^{\mu}{}_{\hat 0} &=\gamma (1,-v\sin\varphi, v\cos \varphi, 0)\,,\\
\label{I6}\lambda^{\mu}{}_{\hat 1}&=(0,\cos \varphi, \sin \varphi, 0)\,,\\
\label{I7}\lambda^{\mu}{}_{\hat 2}&=\gamma (v,-\sin \varphi, \cos \varphi, 0)\,,\\
\label{I8}\lambda^{\mu}{}_{\hat 3}&=(0, 0, 0, 1)\,.
\end{align}
Let us note that the unit axes of the observer's spatial frame $\lambda^{\mu}{}_{\hat i}$, for $i = 1, 2, 3$, point in the radial, tangential and $z$ directions, respectively,  in the cylindrical system of coordinates in the background inertial frame.

The derivation of Eq.~\eqref{I4} is based on the application of the hypothesis of locality to the radiation field together with the nonlocal operation of Fourier analysis in time. Indeed, the measured frequency ${\tilde \omega}_{\pm}$ approaches the local Doppler result $\omega_{\rm D}$ as $\Omega_0/\omega_0 \to 0$, since
\begin{equation}\label{I9}
{\tilde \omega}_{\pm} = \omega_{\rm D}\,\left(1 \mp \frac{\Omega_0}{\omega_0}\right)\,.
\end{equation}
Let us note that $\Omega_0 / \omega_0$ is equal to the ratio of the reduced wavelength of the incident radiation to $c/\Omega_0$, which is the acceleration length of the rotating observer. Therefore, it follows from Eq.~\eqref{I9} that we recover the pointwise transverse Doppler result in the eikonal (or WKB) limit; moreover, phase invariance in inertial frames does not in general extend to accelerated systems. 

Ignoring time dilation for the moment, ${\tilde \omega}_{\pm} \approx \omega_0 \mp \Omega_0$ comes about as a consequence of the coupling of the helicity of the radiation with the rotation of the observer. The incident radiation can be expressed in the circular polarization basis as a linear superposition of positive and negative helicity components. In the positive (negative) helicity case, static inertial observers in the background global inertial frame observe the electric and magnetic radiation fields rotate in the positive (negative) sense with frequency $\omega_0$ about the direction of propagation of the wave. From the viewpoint of the rotating observer, the electric and magnetic radiation fields rotate with frequency $\omega_0 - \Omega_0$ ($\omega_0 + \Omega_0$)
 in the positive (negative) helicity case. This ``angular Doppler effect" has been studied in classical optics; see~\cite{BM4} and the references cited therein.  The helicity-rotation coupling is responsible for the phenomenon of phase wrap-up in the GPS~\cite{Ash}. It is also the origin of a certain frequency shift that occurs when circularly polarized radiation passes through a rotating spin flipper; indeed,  for $\Omega_0 \ll \omega_0$, this phenomenon has been observationally known for a long time~\cite{PJA, MNHS}. 

Imagine a uniformly rotating half-wave plate (HWP) instead of the rotating observer in our thought experiment. At the lower edge of the plate the frequency of an incident positive-helicity wave is 
$\omega' \approx \omega_0 - \Omega_0$. Inside the plate, the frequency remains constant and equal to $\omega'$ as the rotation of the plate is assumed to be uniform. At the upper edge of the plate, the connection of $\omega'$ with the frequency $\omega_1$  of the emerging negative-helicity wave is $\omega' \approx \omega_1 + \Omega_0$, so that $\omega_1 - \omega_0 \approx -2\,\Omega_0$. There is therefore a downshift in the frequency of the radiation as it passes through the spin flipper. A corresponding upshift would occur for incident negative-helicity radiation. The general formula for the energy shift is $\mp 2 \hbar\,s\,\Omega_0$, where $s$ is the spin of the radiation field~\cite{MK}. Further discussion of helicity-rotation coupling for photons is contained in Refs.~\cite{Mash, AnMa, HaMa, BMash, Bliokh:2015yhi}. The spin-rotation coupling is a general phenomenon  that is a manifestation of the inertia of intrinsic spin and has extensive observational support; in fact, it has been recently observed directly in neutron interferometry~\cite{BM5, HN, BM5a, SoTi, LR, SP, Pa, ShHe, LaPa, Ran, Pan:2011zza, Arminjon:2013kxa, Werner, HRW, DSH, DDSH, DDKWLSH, MISM, MISM2, MIHSM, MIM, CB, IMM, PAP, KYM, KMN, NaTa, KMMN}. 

Let us now return to Eq.~\eqref{I4} and note that this equation has observational support only for $\Omega_0 \ll \omega_0$. Are there higher-order terms in Eq.~\eqref{I4}? For instance, could there be terms of order $(\Omega_0/\omega_0)^2$ or higher in the wave expansion~\eqref{I9}? The existence of such terms is quite unlikely; otherwise, there would be divergence as $\omega_0 \to 0$. Various aspects of such problems have been discussed in Ref.~\cite{BM3}. There is, however, one important issue that we need to address, namely, the possibility that for incident 
positive-helicity radiation of frequency $\omega_0 = \Omega_0$, we have ${\tilde \omega}_{+} = 0$. By a mere rotation of angular velocity $\Omega_0 = \omega_0$, the whole radiation field stands completely still with respect to the rotating observer. This is not possible for inertial observers in Minkowski spacetime, since  $\omega = 0$ implies $\omega_0 = 0$ in Eq.~\eqref{I1}. However, before the advent of special relativity theory, an observer could in principle move at the speed of light and hence stay completely at rest with a radiation field. The situation here is therefore reminiscent of the pre-relativistic Doppler effect; indeed, Einstein mentioned such a puzzling possibility in connection with the pre-relativistic Doppler formula in his autobiographical notes~\cite{Sch}. This conceptual difficulty was removed for inertial observers via the Lorentz invariance of physical laws. We do the same via an appropriate nonlocal extension of the locality hypothesis  of the standard special relativity theory. 

 The hypothesis of locality, as applied to the electromagnetic radiation field, is the origin of the circumstance that a radiation field can in principle stand completely still with respect to a rotating observer. The locality postulate in turn originates from classical mechanics where the state of a classical particle is determined by its position and velocity. On the other hand, in accordance with the Huygens principle, wave phenomena are in general nonlocal. In 1993, a nonlocal theory of accelerated systems was put forward in which the past history of accelerated observers was taken into account via kernels that would be so constructed as to mitigate the conceptual difficulty of radiation standing completely still~\cite{BM6}.  Using electrons in electromagnetic fields as accelerated systems in the correspondence limit, we have shown that the nonlocal theory is in better agreement with quantum mechanics than the standard theory based on the locality postulate~\cite{BM6a}.  We now turn to a brief description of the theory of acceleration-induced nonlocality, which leads to nonlocal special relativity theory~\cite{BM7, BF}. 

\subsection{Nonlocality of Accelerated Systems}

Imagine an accelerated observer following a path $x^\mu (\tau)$ in Minkowski spacetime in the presence of a field $\psi(x)$, where $x$ here represents an event $x^\mu$. What is the field that the accelerated observer measures?  According to the locality postulate of the standard relativity theory, Lorentz invariance can be applied point by point along the path; in this way, the observer measures a field 
\begin{equation}\label{I10}
{\tilde \psi}(\tau) = \Lambda (\tau)\, \psi (x (\tau))\,,
\end{equation}
where $\Lambda$ is a matrix representation of the Lorentz group. On the other hand, according to Bohr and Rosenfeld, it is not possible to measure even the classical electromagnetic field at a spacetime event by inertial observers; indeed, a spacetime averaging procedure is necessarily involved in such a measurement process~\cite{BR1, BR2}. 

Following Bohr and Rosenfeld~\cite{BR1}, consider for the sake of illustration an extended body of volume $V$ and uniform charge density $\bar{\rho}$ that is placed in an external electric field 
$\mathbf{E}(t, \mathbf{x})$. The Lorentz force law is indispensable for the measurement of the electromagnetic field; indeed, the electric and magnetic fields are defined via the Lorentz force law. In this case, we have
\begin{equation}\label{I11}
\frac{d\mathbf{P}}{dt}  = \bar{\rho}\, \int_V \mathbf{E}(t, \mathbf{x})\,d^3x\,.
\end{equation}
Measuring the momentum of the extended charged body $\mathbf{P}$ at $t'$ and $t'' > t'$, we get
\begin{equation}\label{I12}
\mathbf{P}(t'') - \mathbf{P}(t')  = \bar{\rho}\, \int_{t'}^{t''}\int_V \mathbf{E}(t, \mathbf{x})\,d^3x\,;
\end{equation}
thus, the field as measured by static inertial observers is given by
\begin{equation}\label{I13}
< \mathbf{E} > ~ = \frac{1}{\Delta} \int_\Delta \mathbf{E}(x)\,d^4x\,,
\end{equation}
where $\Delta = (t'' - t') \,V$.

Extending the Bohr--Rosenfeld argument to accelerated systems is a daunting task. Instead, we implement the averaging requirement as follows: Let $\Psi(\tau)$ be the field actually measured by the accelerated observer; then, the most general linear relationship linking $\Psi(\tau)$ to ${\tilde \psi}(\tau) = \Lambda (\tau)\, \psi (x (\tau))$ that is consistent with causality is given by
\begin{equation}\label{I14}
\Psi(\tau) = {\tilde \psi}(\tau) +\int_{\tau_0}^{\tau} \mathcal{K}(\tau, \tau')\,{\tilde \psi}(\tau')\, d\tau'\,
\end{equation}
for $\tau \ge \tau_0$, where $\tau_0$ is the instant at which the acceleration is turned on. In this ansatz, which is consistent with the superposition principle, the locality postulate is extended by a certain average over the past world line of the observer. 

The Volterra integral equation of the second kind~\eqref{I14} relates $\psi$ to $\Psi$ via ${\tilde \psi}$ given by Eq.~\eqref{I10}; indeed, the resulting connection is unique in the space of continuous functions by Volterra's theorem, which has been extended to the space of square-integrable functions by Tricomi~\cite{Volt, Tric}. The Volterra--Tricomi uniqueness result is an important feature of the nonlocal theory of accelerated systems. 

The kernel of ansatz~\eqref{I14}, $\mathcal{K}(\tau, \tau')$, is expected to vanish in the absence of acceleration; moreover, the kernel should be such that a basic radiation field could never stand completely still with respect to an accelerated observer. Detailed investigations have revealed that a consistent theory can be developed based on 
\begin{equation}\label{I15}
\mathcal{K}(\tau ,\tau') = K(\tau')=-\frac{d\Lambda (\tau')}{d\tau'}\, \Lambda^{-1}(\tau');
\end{equation}
see Ref.~\cite{BM7} and the references cited therein. The formula for $K(\tau)$, which vanishes in the absence of the acceleration of the observer, simply follows from the assumption that a constant field $\psi$ will lead to a constant $\Psi$ in ansatz~\eqref{I14}. The Volterra--Tricomi uniqueness theorem then implies that an actual radiation field will never be constant as measured by accelerated observers. Moreover, after the acceleration has been turned off at $\tau_f$, the memory of past acceleration remains as a \emph{constant} addition to the measured field; that is, for $\tau \ge \tau_f$, 
\begin{equation}\label{I16}
\Psi(\tau) = {\tilde \psi}(\tau) +\int_{\tau_0}^{\tau_f} K(\tau')\,{\tilde \psi}(\tau')\, d\tau'\,.
\end{equation}

The nonlocal theory of accelerated systems has been applied to the Dirac field with satisfactory results~\cite{BM8}. There is a problem, however, in the case of the electromagnetic field $F_{\mu \nu}$ due to the existence of electrostatic and magnetostatic fields.  To see this, let us note that with the substitution of kernel~\eqref{I15}, ansatz~\eqref{I14} can be written as
\begin{equation}\label{I17}
\Psi(\tau) = {\tilde \psi}(\tau_0) +\int_{\tau_0}^{\tau} \Lambda(\tau')\,\frac{d\psi(\tau')}{d\tau'}\, d\tau'\,.
\end{equation}
It follows that in a static background electromagnetic field, an accelerated observer will always measure a constant field ${\tilde \psi}(\tau_0)$, which contradicts experimental results involving small accelerations that appear to be in general agreement with the locality postulate of the special relativity theory~\cite{Ken, Peg, Swa}. 

To resolve this difficulty, one can assume that kernel $K$ given by Eq.~\eqref{I15} is appropriate for the vector potential $A_\mu$, while a different kernel should be chosen for the field tensor $F_{\mu \nu}$.  A specific choice for such a kernel involving parity violation has been put forward in Ref.~\cite{BM9}. The purpose of the present paper is to re-examine our previous proposal and provide a deeper and more general analysis of the problem. 

\section{Nonlocal Electrodynamics}

According to the hypothesis of locality, the vector potential $A_\mu$ and the field tensor $F_{\mu \nu}$ as measured by the accelerated observer are
\begin{equation}\label{E1}
\tilde{A}_{\hat \alpha} =A_\mu \,\lambda^{\mu}{}_{\hat \alpha}\,, \qquad \tilde{F}_{\hat \alpha \hat \beta} = F_{\mu\nu}\,\lambda^{\mu}{}_{\hat \alpha}\,\lambda^{\nu}{}_{\hat \beta}\,.
\end{equation}
On the other hand, we claim that the actually measured fields after the acceleration is turned on at $\tau_0$ are
\begin{align}\label{E2}
\mathcal{A}_{\hat \alpha}(\tau )&=\tilde{A}_{\hat \alpha}(\tau)+\int_{\tau_0}^{\tau} K_{\hat \alpha}{}^{\hat \beta}(\tau')\,\tilde{A}_{\hat \beta}(\tau')\,d\tau'\,,\\
\label{E3} \mathcal{F}^{\hat \alpha \hat \beta}(\tau)&=\tilde{F}^{\hat \alpha \hat \beta}(\tau)+\int_{\tau_0}^{\tau} K^{\hat \alpha \hat \beta}{}_{\hat \gamma \hat \delta}(\tau')\,
\tilde{F}^{\hat \gamma \hat \delta}(\tau')\,d\tau'\,.
\end{align}
Once the kernels are fixed, other aspects of the nonlocal electrodynamics of accelerated systems can be determined. For the kernel of Eq.~\eqref{E2} involving the vector potential $A_\mu$, the simplest possibility would be to choose the kernel given by Eq.~\eqref{I15}. If $A_\mu$ is constant, then $F_{\mu \nu} = 0$ everywhere and via Eq.~\eqref{I17} we have that $\mathcal{A}_{\hat \alpha}(\tau )=\tilde{A}_{\hat \alpha}(\tau_0)$ is constant as well. This circumstance does not appear to pose any difficulty; moreover, it is clear that under the gauge transformation $A_\mu \mapsto A_\mu + \partial_\mu \Sigma$,  $\tilde{A}_{\hat \alpha} \mapsto \tilde{A}_{\hat \alpha} 
+ (\partial_\mu \Sigma)\,\lambda^{\mu}{}_{\hat \alpha}$ and, similarly, $\mathcal{A}_{\hat \alpha}$ will be subject to a nonlocal gauge transformation, but Eq.~\eqref{E3} will remain invariant. 

The kernels depend upon the acceleration of the observer; therefore, a description of the acceleration tensor is required.

\subsection{Acceleration Tensor}

Along the observer's world line, the motion of the adapted tetrad frame can be expressed as
\begin{equation}\label{E4}
\frac{d\lambda^{\mu}{}_{\hat \alpha}}{d\tau} = \phi_{\hat \alpha}{}^{\hat \beta}\,\lambda^{\mu}{}_{\hat \beta}\,, \qquad \phi_{\hat \alpha \hat \beta} = - \phi_{\hat \beta \hat \alpha}\,.
\end{equation}
The antisymmetry of the acceleration tensor $\phi_{\hat \alpha \hat \beta}$ is a simple consequence of the orthonormality of the adapted tetrad frame. 

In general, the acceleration tensor has a decomposition, $\phi_{\hat \alpha \hat \beta} \mapsto (-\mathbf{g}, \boldsymbol{\Omega})$, in close analogy with the electromagnetic field tensor, namely, $F_{\mu \nu} \mapsto (\mathbf{E}, \mathbf{B})$, where $F_{0i} = - E_i$ and $F_{ij} = \epsilon_{ijk}\,B^k$ in our convention. The spacetime invariants $\mathbf{g}$ and $\boldsymbol{\Omega}$ have local components with respect to the adapted frame. The ``electric" part of the acceleration tensor corresponds to the translational acceleration of the observer, $g_{\hat i} = \phi_{\hat 0 \hat i}$, while the ``magnetic" part, $\Omega_{\hat i} = \frac{1}{2} \epsilon_{\hat i \hat j \hat k}\, \phi^{\hat j \hat k}$,  corresponds to the angular velocity of the spatial frame of the observer relative to a nonrotating frame that is Fermi-Walker transported along the observer's world line.

\subsection{Acceleration Kernels}

It follows from inspection of Eqs.~\eqref{I10} and~\eqref{I15} as well  as~\eqref{E1} and~\eqref{E4} that for the kernel of the vector potential we have
\begin{equation}\label{E5}
K_{\hat \alpha}{}^{\hat \beta} = - \phi_{\hat \alpha}{}^{\hat \beta}\,,
\end{equation}
which we must use in Eq.~\eqref{E2}.

Let us next turn to the evaluation of the field kernel in Eq.~\eqref{E3}, $K^{\hat \alpha \hat \beta}{}_{\hat \gamma \hat \delta}$, which is antisymmetric in its first and second pairs of indices and is the main subject of the present paper.  It should be constructed out of the Minkowski metric tensor $\eta_{\hat \alpha \hat \beta}$, the Levi-Civita tensor $\epsilon_{\hat \alpha \hat \beta \hat \gamma \hat \delta}$ with 
$\epsilon_{\hat 0 \hat 1 \hat 2 \hat 3} = 1$ and the acceleration tensor $\phi_{\hat \alpha \hat \beta}$. In this connection, the field kernel corresponding to Eq.~\eqref{I15} is given by
\begin{equation}\label{E6} 
\kappa_{\hat \alpha \hat \beta \hat \gamma \hat \delta} = -\frac{1}{2} (\phi_{\hat \alpha \hat \gamma}\, \eta_{\hat \beta \hat \delta} + \phi_{\hat \beta \hat \delta}\, \eta_{\hat \alpha \hat \gamma}
- \phi_{\hat \beta \hat \gamma}\, \eta_{\hat \alpha \hat \delta} - \phi_{\hat \alpha \hat \delta}\, \eta_{\hat \beta \hat \gamma})\,,
\end{equation}
which, as explained before, cannot be the electromagnetic field kernel by itself. Indeed, in previous work~\cite{BM9}, we assumed
\begin{equation}\label{E7} 
K_{\hat \alpha \hat \beta \hat \gamma \hat \delta} = p\, \kappa_{\hat \alpha \hat \beta \hat \gamma \hat \delta} + q\, ^*\kappa_{\hat \alpha \hat \beta \hat \gamma \hat \delta}\,, 
\end{equation}
where $p$ and $q$ are constant real dimensionless parameters and $^*\kappa$ is the dual of $\kappa$, 
\begin{equation}\label{E8} 
^*\kappa^{\hat \alpha \hat \beta}{}_ {\hat \gamma \hat \delta} = \frac{1}{2}\, \epsilon^{\hat \alpha \hat \beta}{}_ {\hat \rho \hat \sigma}\, \kappa^{\hat \rho \hat \sigma}{}_{\hat \gamma \hat \delta}\,. 
\end{equation}

The right and left duals of $\kappa$ turn out to be equal as a consequence of the antisymmetry of the acceleration tensor. Clearly, the existence of $q \ne 0$ violates parity conservation. On the other hand, the adoption of kernel~\eqref{E7} was based on the additional assumption that $K$ is linearly dependent upon the acceleration tensor $\phi_{\hat \alpha \hat \beta}$. 

Let us observe that acceleration-induced nonlocal electrodynamics shares similarities with the nonlocal electrodynamics of media~\cite{MuHM, HeOb}. Even in the linear electrodynamics of media, there is evidence for parity violation; see Refs.~\cite{Hehl:2007jy, Hehl:2007ut, Hehl:2009eqa, Essin:2008rq} and the references cited therein. However, no corresponding observational evidence regarding accelerated systems is available at present. The main purpose of the present work is to point out that one can remove the linearity constraint on the field kernel and thereby develop an acceleration-induced nonlocal electrodynamics that preserves parity conservation. The mathematical form of the various possible field kernels are explored in the last part of Section IV. 

Specifically, in the present work we assume a field kernel that is quadratic in the acceleration tensor and is of the form
\begin{equation}\label{E9} 
K^{\hat \alpha \hat \beta}{}_ {\hat \gamma \hat \delta} = \mu\, \kappa^{\hat \alpha \hat \beta}{}_ {\hat \rho \hat \sigma}\, \kappa^{\hat \rho \hat \sigma}{}_{\hat \gamma \hat \delta}+ \nu\,
\phi^{\hat \alpha \hat \beta}\,\phi_{\hat \gamma \hat \delta}\,,
\end{equation}
where $\mu$ and $\nu$ are constant real parameters that carry dimensions of length. The new field kernel is symmetric; that is, $K_{\hat \alpha \hat \beta \hat \gamma \hat \delta} = 
K_{\hat \gamma \hat \delta \hat \alpha \hat \beta}$. In this paper, we work out the physical consequences of kernel~\eqref{E9} for the sake of illustration; in principle, a parity conserving kernel could be a sum of linear, quadratic and higher-order terms in the acceleration tensor. 

\section{Parity Violation}

Soon after the discovery of parity violation in the weak interactions, Stueckelberg  pointed out the theoretical possibility of parity violation in gravity due to the existence of the covariantly constant Levi-Civita totally antisymmetric tensor field $\epsilon_{\mu \nu \rho \sigma}$ in general relativity~\cite{Stu}. Indeed, this tensor leads to the definition of dual fields that could help induce parity violation. An instance of this circumstance is given in Eq.~\eqref{E7}, the kernel of previously published nonlocal electrodynamics of accelerated systems~\cite{BM9, BM7, BMB}. Observational consequences of the possible breakdown of parity and time-reversal invariance in the gravitational interaction have been considered by a number of authors; see, for instance, Refs.~\cite{KoOk, LeOk, WiRa, BaMa1, BaMa2, Freidel:2005sn, Ivanov:2016krk, 
Conroy:2019ibo, ZZQ, Iosifidis:2020dck, Obukhov:2020zal} and the references cited therein.
On the other hand, there is no parity violation in standard general relativity. Furthermore, parity conservation is maintained  in the standard electrodynamics of accelerated systems and gravitational fields; see Appendix A for a detailed discussion. 

In acceleration-induced nonlocal electrodynamics, when one assumes that the kernel is only linear in acceleration, the existence of parity violating term in Eq.~\eqref{E7} becomes absolutely necessary; otherwise, there would be inconsistency in the physical results of the theory. A somewhat similar situation occurs in the gravitational context involving the torsion pseudovector in the constitutive relation of nonlocal gravity~\cite{HM1, HM2, BMB}. A brief discussion of parity violation in nonlocal gravity is contained in Section VIII. 

There is no observational data regarding accelerated systems that could be used to exclude completely parity violation in electrodynamics. The main purpose of the present paper is to examine quadratic terms in acceleration that could supplant parity violating terms in the kernel. The rest of this paper is devoted to the derivation of some consequences of parity conserving nonlocal electrodynamics of accelerated systems. To this end, it is convenient to express the main integral relation~\eqref{E3} in matrix form. 

\section{Matrix Representation}

For the sake of simplicity, we wish to represent our main field ansatz as a matrix equation. We use the antisymmetry of the field tensor $F^{\mu \nu}$ and assume that its pair of indices in effect takes the values $\{01, 02, 03, 23, 31,12\}$ and we can then take care of the remaining possibilities by means of a simple factor of $2$. Thus, we replace $F^{\mu \nu}$ by a column 6-vector $F$ with components $\mathbf{E}$ and $\mathbf{B}$, since $E_i = F^{0i}$ and $B_i = \frac{1}{2}\,\epsilon_{ijk}\,F^{jk}$ in our convention,
\begin{equation}\label{E10}
F = \left[
\begin{array}{cc}
\mathbf{E} \cr
\mathbf{B} \cr
\end{array}
\right]\,. 
\end{equation}
Similarly, we replace kernel $K^{\hat \alpha \hat \beta}{}_ {\hat \gamma \hat \delta}$ by a $6\times 6$ matrix $\mathbb{K}$ , where the first and second pair of indices range over the set $\{01, 02, 03, 23, 31,12\}$. Equation~\eqref{E3} then takes the form
\begin{equation}\label{E11}
\mathcal{F}(\tau) = \tilde{F}(\tau) + 2 \int_{\tau_0}^{\tau} \mathbb{K}(\tau')\,\tilde{F}(\tau')\,d\tau'\,,
\end{equation}
where $\tilde{F} = \Lambda\, F$ and $\Lambda$ is a $6\times 6$ matrix that can be determined from the adapted tetrad frame field. 

\subsection{Relations Involving Kernels}

We collect here a number of relations that will be useful in explicit applications. It follows from Eq.~\eqref{E6} that for the special kernel $\kappa$ we have
\begin{equation}\label{E12} 
\kappa_{\hat \alpha \hat \beta \hat \gamma \hat \delta} = - \kappa_{\hat \gamma \hat \delta \hat \alpha \hat \beta}\,,\qquad
\kappa^{\mu}{}_{\hat \alpha \hat \mu \hat \beta} = - \phi_{\hat \alpha \hat \beta}\,,  \qquad \kappa^{\hat \alpha \hat \beta}{}_ {\hat \alpha \hat \beta} = 0\,.
\end{equation}
Moreover, $^*\kappa = \kappa^*$ with
\begin{equation}\label{E13} 
\kappa^*_{\hat \alpha \hat \beta}{}^{\hat \gamma \hat \delta} = \frac{1}{2}\, \kappa_{\hat \alpha \hat \beta}{}^{\hat \rho \hat \sigma}\,\epsilon_{\hat \rho \hat \sigma}{}^{\hat \gamma \hat \delta}\, 
= -\frac{1}{2}\,(\phi_{\hat \alpha}{}^{\hat \rho}\,\epsilon_{\hat \rho \hat \beta}{}^{\hat \gamma \hat \delta}-\phi_{\hat \beta}{}^{\hat \rho}\,\epsilon_{\hat \rho \hat \alpha}{}^{\hat \gamma \hat \delta})\,.
\end{equation}
The mixed dual 
\begin{equation}\label{E14} 
\zeta_{\hat \alpha \hat \beta}{}^{\hat \gamma \hat \delta} = \frac{1}{2}\, \kappa_{\hat \alpha \hat \rho \hat \sigma \hat \beta}\,\epsilon^{\hat \rho \hat \sigma \hat \gamma \hat \delta}\, 
\end{equation}
is symmetric in its first pair of indices 
\begin{equation}\label{E15} 
\zeta_{\hat \alpha \hat \beta}{}^{\hat \gamma \hat \delta} = \frac{1}{4}\,(\kappa_{\hat \alpha \hat \rho \hat \sigma \hat \beta} - \kappa_{\hat \alpha \hat \sigma \hat \rho \hat \beta})\,\epsilon^{\hat \rho \hat \sigma \hat \gamma \hat \delta} = \frac{1}{4}\,(\kappa_{\hat \alpha \hat \rho \hat \sigma \hat \beta} + \kappa_{\hat \beta \hat \rho \hat \sigma \hat \alpha})\,\epsilon^{\hat \rho \hat \sigma \hat \gamma \hat \delta}\,. 
\end{equation}
It is therefore not suitable as a field kernel in any case; here, we correct the erroneous assertion that the mixed dual vanishes~\cite{BM9, BM7}.

In expression~\eqref{E6} for the special kernel $\kappa$, we can replace the acceleration tensor $\phi_{\hat \alpha \hat \beta}$ by its dual  $\phi^*_{\hat \alpha \hat \beta}$ and thereby obtain 
$\chi_{\hat \alpha \hat \beta \hat \gamma \hat \delta}$; then, 
\begin{equation}\label{E16} 
\chi_{\hat \alpha \hat \beta}{}^{\hat \gamma \hat \delta} =  \kappa^*_{\hat \alpha \hat \beta}{}^{\hat \gamma \hat \delta}\,, \qquad \chi^*_{\hat \alpha \hat \beta}{}^{\hat \gamma \hat \delta} = - \kappa_{\hat \alpha \hat \beta}{}^{\hat \gamma \hat \delta}\,, 
\end{equation}
so that the double duality operation equals minus the identity operation in this case. 

Regarding quadratic terms in the acceleration tensor, we recall that 
\begin{equation}\label{E17} 
\phi_{\hat \alpha \hat \beta}\,\phi^{\hat \alpha \hat \beta} = - 2\,(g^2 - \Omega^2)\,, \qquad \phi^*_{\hat \alpha \hat \beta}\,\phi^{\hat \alpha \hat \beta} = -2\, \mathbf{g} \cdot \boldsymbol{\Omega}\,.
\end{equation}
Using a function $W$ of these invariants, we can form field kernels of the form $W\,\epsilon^{\hat \alpha \hat \beta}{}_{\hat \gamma \hat \delta}$.  Moreover, we find
\begin{equation}\label{E18} 
\kappa^{\hat \alpha \hat \beta}{}_{\hat \gamma \hat \delta} \, \kappa^{\hat \gamma \hat \delta}{}_{\hat \rho \hat \sigma} = 
\phi^{\hat \alpha}{}_{\hat \rho}\,\phi^{\hat \beta}{}_{\hat \sigma} - \phi^{\hat \alpha}{}_{\hat \sigma}\,\phi^{\hat \beta}{}_{\hat \rho} + \frac{1}{2}\,(\phi^{\hat \alpha}{}_{\hat \gamma}\,\phi^{\hat \gamma}{}_{\hat \rho}\,\delta^{\hat \beta}_{\hat \sigma} - \phi^{\hat \alpha}{}_{\hat \gamma}\,\phi^{\hat \gamma}{}_{\hat \sigma}\,\delta^{\hat \beta}_{\hat \rho} - \phi^{\hat \beta}{}_{\hat \gamma}\,\phi^{\hat \gamma}{}_{\hat \rho}\,\delta^{\hat \alpha}_{\hat \sigma} + \phi^{\hat \beta}{}_{\hat \gamma}\,\phi^{\hat \gamma}{}_{\hat \sigma}\,\delta^{\hat \alpha}_{\hat \rho})\,
\end{equation}
and
\begin{equation}\label{E19} 
\kappa^{\hat \alpha \hat \beta}{}_{\hat \gamma \hat \delta} \, \kappa^{\hat \gamma \hat \delta}{}_{\hat \alpha \hat \beta} = - 2\,\phi_{\hat \rho \hat \sigma}\,\phi^{\hat \rho \hat \sigma}\,.
\end{equation}

In connection with the matrix representation, a detailed treatment reveals that for the special kernel $\kappa$ we have
\begin{equation}\label{E20} 
\kappa^{\hat \alpha \hat \beta}{}_ {\hat \gamma \hat \delta} \mapsto \kappa = \frac{1}{2}
\begin{bmatrix} 
\kappa_1 & -\kappa_2\\
\kappa_2 &\kappa_1 
\end{bmatrix}\,, 
\end{equation}
where the $3\times 3$ matrices $\kappa_1$ and $\kappa_2$ are given by
\begin{equation}\label{E21}
 \kappa_1 =\boldsymbol{\Omega} \cdot \mathbf{I}\,, \qquad \kappa_2 =\mathbf{g}\cdot \mathbf{I}\,.
\end{equation} 
Here, $I_i$ is  proportional to the operator of infinitesimal rotations about the $x^i$ axis, so that $I_1$, $I_2$ and $I_3$ form a basis for the Lie algebra of the rotation group as a 3-dimensional vector space; that is, 
\begin{equation}\label{E22} 
(I_i)_{jk} =-\epsilon _{ijk}\,, \qquad  [I_i, I_j] = \epsilon _{ijk}\,I_k\,.
\end{equation}
Let us note that in our matrix representation
\begin{equation}\label{E23} 
\epsilon^{\hat \alpha \hat \beta}{}_ {\hat \gamma \hat \delta} \mapsto 
\begin{bmatrix} 
0 & - I\\
I & 0 
\end{bmatrix}\,, 
\end{equation}
where $I =\,$diag$(1, 1, 1)$ is the identity matrix; therefore, Eq.~\eqref{E8} results in
\begin{equation}\label{E24} 
^*\kappa^{\hat \alpha \hat \beta}{}_ {\hat \gamma \hat \delta} \mapsto {^\ast\kappa} = \frac{1}{2}
\begin{bmatrix} 
-\kappa_2 & -\kappa_1\\
\kappa_1 & -\kappa_2 
\end{bmatrix}\,.
\end{equation}

Using the matrix representation, it is straightforward to show that   
\begin{equation}\label{E25} 
\kappa^{\hat \alpha \hat \beta}{}_ {\hat \rho \hat \sigma}\, \kappa^{\hat \rho \hat \sigma}{}_{\hat \gamma \hat \delta} = - ^*\kappa^{\hat \alpha \hat \beta}{}_ {\hat \rho \hat \sigma}\, ^*\kappa^{\hat \rho \hat \sigma}{}_{\hat \gamma \hat \delta}\,, \quad ^*\kappa^{\hat \alpha \hat \beta}{}_ {\hat \rho \hat \sigma}\, \kappa^{\hat \rho \hat \sigma}{}_{\hat \gamma \hat \delta} = \kappa^{\hat \alpha \hat \beta}{}_ {\hat \rho \hat \sigma}\, ^*\kappa^{\hat \rho \hat \sigma}{}_{\hat \gamma \hat \delta}\,.
\end{equation}
Furthermore, 
\begin{equation}\label{E26} 
\kappa^{\hat \alpha \hat \beta}{}_ {\hat \rho \hat \sigma}\, \kappa^{\hat \rho \hat \sigma}{}_{\hat \gamma \hat \delta} \mapsto \frac{1}{2}
\begin{bmatrix} 
\kappa_1^2-\kappa_2^2 & -(\kappa_1\,\kappa_2 +\kappa_2\,\kappa_1)\\
\kappa_1\,\kappa_2 +\kappa_2\,\kappa_1 & \kappa_1^2-\kappa_2^2 
\end{bmatrix}\,,
\end{equation}
where 
\begin{equation}\label{E27}
\mathrm{tr}(\kappa_1^2) = - 2\, \Omega^2 \,, \qquad \mathrm{tr}(\kappa_2^2) = -2\,g^2\,.
\end{equation} 
For the other quadratic kernel in Eq.~\eqref{E9}, we find the $6\times 6$ representation
\begin{equation}\label{E28} 
\phi^{\hat \alpha \hat \beta}\,\phi_{\hat \gamma \hat \delta} \mapsto
\begin{bmatrix} 
- g_i\,g_j & - g_i\,\Omega_j \\
\Omega_i\,g_j & \Omega_i\,\Omega_j 
\end{bmatrix}\,;
\end{equation}
indeed, similar expressions exist for $^*\phi^{\hat \alpha \hat \beta}\,\phi_{\hat \gamma \hat \delta}$ as well as $\phi^{\hat \alpha \hat \beta}\,^*\phi_{\hat \gamma \hat \delta}$. 

Assuming parity conservation, our quadratic kernel $\mathbb{K}$ given by Eq.~\eqref{E9} has a matrix representation that is given by $\mu$ times the matrix in Eq.~\eqref{E26} plus $\nu$ times the matrix in Eq.~\eqref{E28}; that is,
\begin{equation}\label{E29} 
\mathbb{K} = \frac{\mu}{2}\,
\begin{bmatrix} 
\kappa_1^2-\kappa_2^2 & -(\kappa_1\,\kappa_2 +\kappa_2\,\kappa_1)\\
\kappa_1\,\kappa_2 +\kappa_2\,\kappa_1 & \kappa_1^2-\kappa_2^2 
\end{bmatrix}
+ \nu\,
\begin{bmatrix} 
- g_i\,g_j & - g_i\,\Omega_j \\
\Omega_i\,g_j & \Omega_i\,\Omega_j 
\end{bmatrix}\,.
\end{equation}

 We examine the physical consequences of the quadratic kernel~\eqref{E9} in the next three sections. In Section V, we use the matrix formalism to explore helicity-rotation coupling, while in Sections VI and VII we study the case of constant background electromagnetic fields. We will find that of the two terms in our proposed quadratic kernel~\eqref{E29}, only the first term gives reasonable results, while the second term must vanish (i.e., $\nu = 0$). Section VIII contains a general discussion of our results.

\section{Influence of the Quadratic Kernel on Helicity-Rotation Coupling}

The purpose of this section is to examine the implications of  kernel~\eqref{E29} for the phenomenon of spin-rotation coupling. We therefore imagine an incident plane monochromatic electromagnetic wave that is normally incident on the rotating observer under consideration in this work. Equation~\eqref{E11} is linear by assumption; hence, we can employ complex fields in our calculations and assign physical significance to their real parts in our convention. We represent the incident wave in the circular polarization basis by
\begin{equation}\label{R1} 
F_\pm (t,\mathbf{x})=i\,\omega_0\, a_{\pm} 
\begin{bmatrix} \mathbf{e}_\pm\\ \mathbf{b}_\pm
\end{bmatrix} 
e^{-i\,\omega_0\, (t-z)}, 
\end{equation} 
where 
\begin{equation}\label{R2} 
\mathbf{e_\pm} = \frac{1}{\sqrt{2}}\,(\widehat{\mathbf{x}} \pm i\,\widehat{\mathbf{y}})\,, \qquad \mathbf{b}_\pm =\mp\, i\,\mathbf{e}_\pm\,, \qquad \mathbf{e}_\pm \cdot \mathbf{e}^\ast_{\pm}=1\, 
\end{equation} 
and the upper (lower) sign represents positive (negative) helicity radiation. Here, $\widehat{\mathbf{x}}$ and $\widehat{\mathbf{y}}$ are unit vectors in the $x$ and $y$ directions, respectively, while $\omega_0$ is the frequency  and $a_{+}$ ($a_{-}$) is the constant amplitude of positive (negative) helicity incident radiation. 

Next, we imagine the instantaneous projection of this field on the tetrad frame~\eqref{I5}--\eqref{I8} of the rotating observer. The acceleration tensor for the rotating observer has a translational part that corresponds to the centripetal acceleration $\mathbf{g}$ and a rotational part that corresponds to the angular velocity $\boldsymbol{\Omega}$ with respect to a locally nonrotating frame. These have components $\mathbf{g}=-v\,\gamma^2\,\Omega_0(1,0,0)$ and $\boldsymbol{\Omega} =\gamma^2\,\Omega_0 (0,0,1)$ with respect to the local adapted spatial frame $\lambda^{\mu}{} _{\hat i}$, $i=1,2,3$, that indicate the radial, tangential and $z$ directions, respectively. The projected field $\tilde{F}$ is given by $\tilde{F} = \Lambda\,F$, where the $6\times 6$ matrix $\Lambda$ is obtained from Eq.~\eqref{E1} and can be expressed as
\begin{equation}\label{R3} 
\Lambda =  
\begin{bmatrix}
 \Lambda_1 & - \Lambda_2 \\ 
 \Lambda_2 & \Lambda_1
\end{bmatrix} \,, 
\end{equation} 
where
\begin{equation}\label{R4} 
\Lambda_1 =  
\begin{bmatrix}
\gamma \cos \varphi & \gamma \sin \varphi & 0 \\ 
 - \sin \varphi  & \cos \varphi & 0 \\
 0 & 0 & \gamma
\end{bmatrix} 
\,, \qquad \Lambda_2 = v\,\gamma
\begin{bmatrix}
0 & 0 & -1 \\ 
 0  & 0 & 0 \\
\cos \varphi &  \sin \varphi & 0
\end{bmatrix} 
\,.
\end{equation} 
Let us recall here that $\varphi = \Omega_0\,t = \gamma\,\Omega_0 \,\tau$. 

According to the locality postulate, the projected field $\tilde{F}$ measured by the momentarily comoving inertial observers along the world line of the rotating observer is given by
\begin{equation}\label{R5} 
\tilde{F}_\pm (\tau ) = i\,\gamma\, \omega_0\, a_\pm 
\begin{bmatrix} 
\tilde{\mathbf{e}}_\pm \\ \tilde{\mathbf{b}}_\pm 
\end{bmatrix}
e^{-i\,\tilde{\omega}_{\pm} \tau}\,,
\end{equation}
where  
\begin{equation}\label{R6} 
\tilde{\mathbf{e}}_\pm =\frac{1}{\sqrt{2}} 
\begin{bmatrix} 1\\ \pm i\,\gamma^{-1}\\ \pm i\,v 
\end{bmatrix}
\,, \qquad  \tilde{\mathbf{b}}_\pm =\mp\, i\,\tilde{\mathbf{e}}_\pm\,, \qquad \tilde{\mathbf{e}}_\pm \cdot \tilde{\mathbf{e}}_\pm ^\ast=1\,.
\end{equation}
As expected, we recover Eq.~\eqref{I4}, $\tilde{\omega}_{\pm} =\gamma (\omega_0 \mp \Omega_0)$, which is the frequency of the wave as measured by the momentarily tangent inertial observer; furthermore, the relative strength of the measured amplitudes of the helicity states in Eq.~\eqref{R5} is still $(a_{+}/a_{-})$ and is therefore not affected by the observer's rotation.

Finally, we must compute the integral of $\mathbb{K}\,\tilde{F}$ over the proper time of the rotating observer in accordance with Eq.~\eqref{E11}.  The second term in the quadratic kernel~\eqref{E29}, which is represented by matrix~\eqref{E28}, does not contribute to the integral in Eq.~\eqref{E11} because $\phi_{\hat \alpha \hat \beta}\,\tilde{F}^{\hat \alpha \hat \beta} = 0$ in this case based on the circumstance that  $g_1 + v\,\Omega_3 = 0$ for the uniformly rotating observer.  Regarding the first term in Eq.~\eqref{E29}, we have
\begin{equation}\label{R7a}
\kappa_1^2  = - \gamma^4\, \Omega_0^2\,\, \mathrm{diag}(1, 1, 0)\,, \qquad \kappa_2^2 = - v^2\,\gamma^4\, \Omega_0^2\,\, \mathrm{diag}(0, 1, 1)\,.
\end{equation}
Moreover, the elements of the matrix~\eqref{E26} are given by
\begin{equation}\label{R7}
\kappa_1^2 -\kappa_2^2 = - \gamma^4\, \Omega_0^2\,\, \mathrm{diag}(1, \gamma^{-2}, -v^2)\,, \qquad \kappa_1\,\kappa_2 +\kappa_2\,\kappa_1 = - v\,\gamma^4\, \Omega_0^2\, 
\begin{bmatrix}
0 & 0 &1\\ 
0 & 0 & 0 \\
1 & 0 & 0 
\end{bmatrix}\,. 
\end{equation}
It follows from Eq.~\eqref{E11} that for $\tau \ge 0$, we have
\begin{equation}\label{R8}
\mathcal{F}_{\pm}(\tau) = \left[1 + i\,\mu\,\gamma^2 \,\Omega_0^2\,\frac{e^{i\,\tilde{\omega}_{\pm}\,\tau} - 1}{\tilde{\omega}_{\pm}}\right]\,\tilde{F}_{\pm}(\tau)\,.
\end{equation}
It is important to note that $\mathcal{F}_{\pm}(\tau)$ can never become constant; moreover, for incident positive-helicity radiation with resonance frequency $\omega_0 =\Omega_0$, we have 
$\tilde{\omega}_{+} = 0$ and
\begin{equation}\label{R9}
 \mathcal{F}_{+} (\tau )= (1- \mu\,\gamma^2\, \Omega _0^2\,\tau)\,\tilde{F}_{+}\,,
\end{equation}
where $\tilde{F}_{+}$ is constant. In the measured field~\eqref{R8}, the ratio of the positive-helicity wave amplitude to the negative-helicity wave amplitude is given by $(a_{+}/a_{-})\, \rho$, where
\begin{equation}\label{R10} 
\rho = \frac{\omega_0 + \Omega_0}{\omega_0 - \Omega_0} ~ \frac{\omega_0 - \Omega_0- i\, \mu\,\gamma\,\Omega_0^2
}{\omega_0 + \Omega_0- i\, \mu\,\gamma\,\Omega_0^2}\,.
\end{equation}
 If $\mu = 0$, then $\rho = 1$, in accordance with the observation that the locality postulate does not affect the relative strength of the helicity amplitudes~\cite{BM6a, BM7}. However, when $\mu \ne 0$,  $|\rho| > 1$, as expected; that is, for an observer that is rotating uniformly in the positive sense about the $z$ axis, the measured amplitude of the incident positive-helicity radiation is enhanced, while the measured amplitude of the negative-helicity radiation is diminished~\cite{BM6a, BM7, BM9}. This result is independent of the sign and magnitude of $\mu \ne 0$.

\section{Rotating Observer in Constant Electromagnetic Field}

Consider a constant electromagnetic field in the background global inertial frame. Suppose that for $\tau > 0$, the uniformly rotating observer measures the electromagnetic field. Thus in our nonlocal ansatz, Eq.~\eqref{E11}, we use the quadratic kernel~\eqref{E29} together with $\tilde{F} = \Lambda\, F$, where $F$ given in Eq.~\eqref{E10} is constant and $\Lambda$ is given by Eq.~\eqref{R3}. 

For the sake of simplicity, it is useful to define
\begin{equation}\label{S1} 
C := \cos \varphi\,, \qquad S := \sin \varphi\,,\qquad \varphi = \Omega_0\,t = \gamma \,\Omega_0\,\tau\,
\end{equation}
and auxiliary dimensionless parameters $\mu'$ and $\nu'$ given by
\begin{equation}\label{S2} 
\mu' := \gamma^2\, \Omega_0\,\mu\,, \qquad \nu' = 2\,\gamma\,\Omega_0\,\nu\,.
\end{equation}
A detailed but straightforward calculation reveals that  
\begin{equation}\label{S3} 
\mathcal{E}_1 = \gamma\, (C\,E_1 +S\,E_2) - \mu'\, [S\,E_1+(1-C)\,E_2] + v \gamma\, (1+\nu'\,\varphi) B_3\,, 
\end{equation}
\begin{equation}\label{S4} 
\mathcal{E}_2 =  - S\,E_1 +C\,E_2 + \gamma^{-1}\, \mu'\, [(1-C)\,E_1 - S\,E_2]\,, 
\end{equation}
\begin{equation}\label{S5} 
\mathcal{E}_3 = \gamma\,E_3 - v \gamma\, (C\,B_1 +S\,B_2) + v \mu'\, [S\,B_1+(1-C)\,B_2]\,, 
\end{equation}
\begin{equation}\label{S6} 
\mathcal{B}_1 = \gamma\, (C\,B_1 +S\,B_2) -  \mu'\, [S\,B_1+(1-C)\,B_2] -v \gamma\, E_3\,, 
\end{equation}
\begin{equation}\label{S7} 
\mathcal{B}_2 =  - S\,B_1 +C\,B_2 + \gamma^{-1}\, \mu'\, [(1-C)\,B_1 - S\,B_2]\,, 
\end{equation}
\begin{equation}\label{S8} 
\mathcal{B}_3 = v\gamma\, (C\,E_1 +S\,E_2) - v\mu'\, [S\,E_1+(1-C)\,E_2] +  \gamma\, (1+\nu'\,\varphi) B_3\,. 
\end{equation}

It is important to observe that while terms proportional to $\mu'$ are harmonic, terms proportional to $\nu'$ are secular. That is, as the observer rotates uniformly, the $\mu'$ terms are simply periodic with proper period $2\,\pi/(\gamma\,\Omega_0)$, but the $\nu'$ terms are cumulative. If $B_3 \ne 0$, then $\mathcal{E}_1$ and $\mathcal{B}_3$ can grow linearly with time. This is not physically acceptable and indicates that we should eventually drop the second term in the quadratic field kernel (i.e., $\nu = 0$).   

\section{Linearly Accelerated Observer in Constant Field}

Consider next an observer moving with uniform acceleration along the $z$ axis for $t > 0$. For $-\infty < t < 0$, the observer is at rest at the spatial origin of a Cartesian coordinate system $(x, y, z)$ in an inertial reference frame. The acceleration is turned on at $t = \tau_0 = 0$ and will be turned off at $\tau_f > 0$. For $0 \le \tau \le \tau_f$, the tetrad frame of the observer in $(t, x, y, z)$ coordinates is given by 
\begin{align}
\label{L1}\lambda^{\mu}{}_{\hat 0} &=(\mathbb{C}, 0, 0, \mathbb{S})\,,\\
\label{L2}\lambda^{\mu}{}_{\hat 1}&=(0, 1, 0, 0)\,,\\
\label{L3}\lambda^{\mu}{}_{\hat 2}&= (0, 0, 1, 0)\,,\\
\label{L4}\lambda^{\mu}{}_{\hat 3}&=(\mathbb{S}, 0, 0, \mathbb{C})\,,
\end{align}
where 
\begin{equation}\label{L5} 
\mathbb{C} = \cosh (g\,\tau)\,, \qquad \mathbb{S} = \sinh (g\,\tau)\,. 
\end{equation}
Here, $g$ is the constant invariant translational acceleration of the observer; that is, the acceleration tensor in this case is given by $\mathbf{g} = g\,(0, 0,1)$ and $\boldsymbol{\Omega} = 0$ with respect to the local adapted spatial frame $\lambda^{\mu}{} _{\hat i}$, $i=1,2,3$, that indicate the Cartesian axes $x$, $y$ and $z$ directions, respectively.  The observer's Lorentz factor and speed are given by $\mathbb{C}$ and $\mathbb{S}/\mathbb{C}$, respectively.

The background field $F$ given by Eq.~\eqref{E10} is again constant. The field as measured by the instantaneously tangent inertial observer is $\tilde{F} = \Lambda\,F$, where $\Lambda$ has the same form as in Eq.~\eqref{R3}, but with $\Lambda_1$ and $\Lambda_2$ given by
\begin{equation}\label{L6} 
\Lambda_1 = \mathrm{diag} (\mathbb{C}, \mathbb{C}, 1)\,, \qquad \Lambda_2 = -\mathbb{S}\,I_3\,. 
\end{equation}
The special kernel $\kappa$ given by Eq.~\eqref{E20} reduces in this case to 
\begin{equation}\label{L7} 
\kappa_1 = 0\,, \qquad \kappa_2 = g\,I_3\,, \qquad I_3 =
\begin{bmatrix}
0 & -1 &0\\ 
1 & 0 & 0 \\
0 & 0 & 0 
\end{bmatrix}\,. 
\end{equation}
In the formula for $\mathbb{K}$, we need $-\kappa_2^2 =  g^2\, \mathrm{diag}(1, 1, 0)$; hence, 
\begin{equation}\label{L7a} 
2\, \mathbb{K} = g^2\, \mathrm{diag} (\mu, \mu, -2\,\nu, \mu, \mu, 0)\,. 
\end{equation}
As before, a detailed but straightforward calculation based on Eq.~\eqref{E11} reveals that   
\begin{equation}\label{L8} 
\mathcal{E}_1 =  \mathbb{C}\,E_1 - \mathbb{S}\,B_2 + \mu\, g\,[\mathbb{S}\,E_1 -(\mathbb{C} - 1)\,B_2]\,, 
\end{equation}
\begin{equation}\label{L9} 
\mathcal{E}_2 =  \mathbb{C}\,E_2 + \mathbb{S}\,B_1  + \mu\,g\, [\mathbb{S}\,E_2 + (\mathbb{C} - 1)\,B_1]\,, 
\end{equation}
\begin{equation}\label{L10} 
\mathcal{E}_3 = (1-2\, \nu\,g^2 \,\tau)E_3 \,, 
\end{equation}
\begin{equation}\label{L11} 
\mathcal{B}_1 = \mathbb{C}\,B_1 + \mathbb{S}\,E_2 +  \mu\,g\, [\mathbb{S}\,B_1+(\mathbb{C} - 1)\,E_2]\,, 
\end{equation}
\begin{equation}\label{L12} 
\mathcal{B}_2 = \mathbb{C}\,B_2 - \mathbb{S}\,E_1   +  \mu\,g\, [\mathbb{S}\,B_2 - (\mathbb{C}-1)\,E_1]\,, 
\end{equation}
\begin{equation}\label{L13} 
\mathcal{B}_3 =  B_3\,. 
\end{equation}
The only contribution from the second term in the quadratic kernel~\eqref{E29} occurs in $\mathcal{E}_3 = (1-2\, \nu\,g^2 \tau)E_3$, which is again cumulative.  If $E_3 \ne 0$, then the measured magnitude of the electric field along the direction of motion varies linearly with proper time. As before, it seems appropriate to set $\nu = 0$.

\section{Discussion}

It is noteworthy that parity violating terms had to be introduced for the sake of consistency in the theories of acceleration-induced nonlocal electrodynamics as well as nonlocal gravity~\cite{BMB}. On the other hand, it is important to understand theoretically why their presence was deemed necessary; that is, one has to find out whether these were genuine or somehow spurious and a consequence of some assumption that had no firm physical foundation. It is simpler to look at the problem in nonlocal electrodynamics, where the kernel was originally assumed to be linear in the acceleration tensor. Removing this restriction has been the main objective of the present work. We find that the kernel could in principle contain linear,  quadratic and possibly higher-order parity conserving terms in the acceleration tensor. Concentrating on quadratic kernels, we have worked out some of the physical  consequences of an acceleration-induced nonlocal field theory of electrodynamics that involves a
kernel given by
\begin{equation}\label{D1} 
K^{\hat \alpha \hat \beta}{}_ {\hat \gamma \hat \delta} = \mu\, \kappa^{\hat \alpha \hat \beta}{}_ {\hat \rho \hat \sigma}\, \kappa^{\hat \rho \hat \sigma}{}_{\hat \gamma \hat \delta}\,,
\end{equation}
where $\mu$ is a constant parameter with dimensions of length that must be determined from comparison of the theory with observational data. 

What are the implications of the present investigation for the problem of parity violation in nonlocal gravity? Let us note that general relativity (GR) has an equivalent tetrad formulation as a teleparallel theory of gravity (GR$_{||}$).  Nonlocal gravity is indeed the nonlocal version of GR$_{||}$, the teleparallel equivalent of GR. Specifically, consider a tetrad field $e^{\mu}{}_{\hat \alpha}(x)$ adapted to a set of preferred observers in spacetime. The tetrad system is orthonormal, i.e., 
\begin{equation}\label{D2}
g^{\mu \nu} = \eta^{\hat \alpha \hat \beta}\,e^{\mu}{}_{\hat \alpha}\,e^{\nu}{}_{\hat \beta}\,,
\end{equation}
which defines the spacetime metric tensor $g_{\mu \nu}(x)$. The gravitational field in this framework is given by the torsion tensor
\begin{equation}\label{D3}
 C_{\mu \nu}{}^{\hat{\alpha}}= \partial_{\mu}e_{\nu}{}^{\hat{\alpha}}-\partial_{\nu}e_{\mu}{}^{\hat{\alpha}}\,,
\end{equation}
which, for each $\alpha = 0, 1, 2, 3$, bears a strong resemblance to the electromagnetic field tensor. Using the torsion tensor, one can define  three independent algebraic invariants given by
\begin{equation}\label{D4}
\mathcal{I}_1=C_{\alpha \beta \gamma}C^{\alpha \beta \gamma}, \quad \mathcal{I}_2=C_{\alpha \beta \gamma}C^{\gamma \beta \alpha}, \quad \mathcal{I}_3=C_\alpha C^\alpha\,,
\end{equation}
where $C_\alpha$ is the torsion vector: $C_\alpha = C_{\beta \alpha}{}^{\beta} = - C_{\alpha}{}^{\beta}{}_{\beta}$. With a Lagrangian density
\begin{equation}\label{D5}
\mathcal{L}_g = - \frac{c^3}{32\,\pi\,G}\,\sqrt{-g}\,\left(\frac{1}{2}\,\mathcal{I}_1+\mathcal{I}_2-2\,\mathcal{I}_3\right),
\end{equation}
we recover the field equations of GR$_{||}$, which exhibit formal similarities with the Maxwell field equations in a material medium. In the nonlocal electrodynamics of media, nonlocality enters the theory via constitutive relations. The same idea can be implemented in nonlocal gravity; that is, the local constitutive relation of GR$_{||}$ can be rendered nonlocal via the introduction of a scalar kernel. The resulting theory has only one known trivial exact solution, namely, the Minkowski spacetime in the absence of gravitation. It has not been possible thus far to find any nontrivial exact solution of the theory~\cite{Bini:2016phe}; instead, linearized GR$_{||}$ has been studied in detail~\cite{BMB}.  As elucidated in Ref.~\cite{BMB}, the equations of linearized nonlocal gravity exhibit a certain inconsistency that can be eliminated by adding to the constitutive relation a parity violating term proportional to
\begin{equation}\label{D6}
\check{C}_\mu\, g_{\nu \rho}-\check{C}_\nu\, g_{\mu \rho}\,,
\end{equation}
where $\check{C}_\mu$ is the torsion pseudovector given by
\begin{equation}\label{D7}
\check{C}_\mu=\frac{1}{3!} \,C^{\alpha \beta \gamma}\,\epsilon_{\alpha \beta \gamma \mu}\,.
\end{equation}
General discussions of parity violation in tetrad theories of gravitation are contained in Refs.~\cite{Baekler:2011jt, Itin:2018dru}. In connection with nonlocal GR$_{||}$, various theoretical aspects of parity violating term~\eqref{D6} have been explored in Ref.~\cite{Itin:2018dru}.  
Can one avoid parity violation in nonlocal gravity as in the nonlocal electrodynamics of accelerated systems?  The resolution of this issue requires further investigation.

\appendix

\section{$P$ and $T$ Invariance in Standard Electrodynamics of Accelerated Systems and Gravitational Fields}

It can be shown that the electromagnetic field equations in an accelerated system or a gravitational field can be reformulated as Maxwell's equations in flat Minkowski spacetime in Cartesian coordinates but in the presence of a certain hypothetical ``medium". To this end, imagine Maxwell's equations
\begin{equation}\label{A1}
F_{[\mu \nu , \rho]} = 0\,, \qquad  (\sqrt{-g}\,F^{\mu \nu})_{,\nu} = 4\,\pi\, j^\mu\,
\end{equation}
on a background spacetime metric $-ds^2 = g_{\mu \nu}\,dx^\mu\,dx^\nu$ expressed in standard Cartesian coordinates $x^\mu = (t, x, y, z)$. We use the natural decompositions
\begin{equation}\label{A2}
F_{\mu \nu} \mapsto (\mathbf{E}, \mathbf{B})\,, \qquad  \sqrt{-g}\,F^{\mu \nu} \mapsto (-\mathbf{D}, \mathbf{H})\,,
\end{equation}
where $\sqrt{-g}\,F^{0i} = D_i$ and $\sqrt{-g}\,F^{ij} = \epsilon^{ijk} H_k$ in our convention. In this way, we recover the original form of Maxwell's equations in Minkowski spacetime but in a background gyrotropic medium with constitutive relations~\cite{Sk, Pl, Fe, VIS}
\begin{equation}\label{A3}
D_i = \epsilon_{ij}\,E_j - (\mathbf{G} \times \mathbf{H})_i\,, \qquad B_i = \mu_{ij}\,H_j + (\mathbf{G} \times \mathbf{E})_i\,,
\end{equation} 
where the conformally invariant properties of the optical medium are given by equal dielectric and permeability tensors as well as a gyration vector which has the interpretation of the gravitational vector potential,
\begin{equation}\label{A4}
\epsilon_{ij} = \mu_{ij} = -\sqrt{-g}\,\frac{g^{ij}}{g_{00}}\,, \qquad G_i = - \frac{g_{0i}}{g_{00}}\,.
\end{equation} 
We briefly discuss the derivation of the constitutive relations~\eqref{A3} at the end of this appendix. There is no \emph{double refraction} when the electric permittivity and magnetic permeability tensors coincide, which is a special property of the medium that simulates acceleration and gravitation in this case~\cite{VIS}.

Under the parity transformation $P: (x, y, z) \mapsto (-x, -y, -z)$ or time reversal $T: t \mapsto -t$,  the background spacetime interval remains invariant, while the components of the metric tensor transform as $(g_{00}, g_{0i}, g_{ij}) \mapsto (g_{00}, - g_{0i}, g_{ij})$. Similarly, we find for the inverse metric $(g^{00}, g^{0i}, g^{ij}) \mapsto (g^{00}, - g^{0i}, g^{ij})$. Furthermore, we can write
\begin{equation}\label{A5}
- ds^2 = g_{00}\,(dt - G_k\,dx^k)^2 + \gamma_{ij}\,dx^i\,dx^j\,,        
\end{equation}
where $(\gamma_{ij})$, defined by
\begin{equation}\label{A6}
\gamma_{ij} := g_{ij} - \frac{g_{0i}\,g_{0j}}{g_{00}}\,,           
\end{equation}
is the inverse of $(g^{ij})$; that is, $g^{ik}\,\gamma_{kj} = \delta^i_j$.
For the determinant of the metric tensor $g := \det(g_{\mu\nu})$, we have the general relation $ g = g_{00}\,\det(\gamma_{ij})$. Moreover, 
\begin{equation}\label{A7}
g_{00} = g\,\det(g^{ij})\,, \qquad g^{00} = \frac{\det(g_{ij})}{g}\,,
\end{equation}   
which imply that $g$ does not change under parity and time reversal. These results are all consistent with the fact that under parity and time reversal,  we have $(\mu_{ij}, \mathbf{G}_i) \mapsto (\mu_{ij}, - \mathbf{G}_i)$. Maxwell's equations in an inertial frame in Minkowski spacetime are such that under parity $\mathbf{E}$ and $\mathbf{D}$ transform as vectors, while $\mathbf{B}$ and $\mathbf{H}$ transform as pseudovectors. Moreover, under time reversal we have $(\mathbf{E}, \mathbf{B}) \mapsto (\mathbf{E}, -\mathbf{B})$ and $(\mathbf{D}, \mathbf{H})\mapsto (\mathbf{D}, -\mathbf{H})$. The constitutive relations~\eqref{A3} remain unchanged separately under these parity and time reversal transformations; therefore, parity is conserved and time reversal invariance is maintained in the standard electrodynamics of accelerated systems and gravitational fields. 

It is interesting to write the general Maxwell equations in Dirac form. For this purpose, we  introduce the Riemann-Silberstein vectors
\begin{equation}\label{A8}
\mathbf{F}^{\pm} = \mathbf{E} \pm i\,\mathbf{H}\,, \qquad  \mathbf{S}^{\pm} = \mathbf{D} \pm i\,\mathbf{B}\,
\end{equation} 
in terms of complex fields, so that Maxwell's equations can now be written as~\cite{BMa, BMb, BMc, BMd} 
\begin{equation}\label{A9}
\frac{1}{i}\,\nabla \times \mathbf{F}^{\pm} = \pm\,\frac{\partial \mathbf{S}^{\pm}}{\partial t} \pm 4\,\pi\,\mathbf{J}\,, \qquad \nabla \cdot \mathbf{S}^{\pm} = 4\,\pi\,\rho\,, 
\end{equation} 
where $j^\mu = (\rho, \mathbf{J})$ and
\begin{equation}\label{A10}
S^{\pm}_p = \mu_{pq}\,F^{\pm}_q  \pm i\, (\mathbf{G} \times \mathbf{F^{\pm}})_p\,. 
\end{equation} 
This formalism has been explicitly employed in Ref.~\cite{HaMa} to study electromagnetic wave propagation in a rotating frame of reference. 

The parity operation can also be applied to the Dirac form of Maxwell's equations given by Eqs.~\eqref{A9} and~\eqref{A10}. For this purpose, we must think of $\pm \,i\,\mathbf{B}$ and $\pm \,i\,\mathbf{H}$ as  vectors and $- \,i\,\mathbf{E}$ as a pseudovector even though $\mathbf{B}$ and $\mathbf{H}$ are axial vectors and $\mathbf{E}$ is a vector. Similarly, under time reversal, we must assume that  $\pm \,i\,\mathbf{B}$ and $\pm \,i\,\mathbf{H}$ remain invariant, while $- \,i\,\mathbf{E}$ changes sign. In this way, the fields become consistent with sources under parity and time reversal operations; that is, we have  $\rho \mapsto \rho$ and $J \mapsto -J$ separately under $P$ and $T$.

An interesting consequence of $\epsilon_{ij} = \mu_{ij}$ should be noted here: Under duality rotations with constant $\alpha$, namely, 
\begin{equation}\label{A11}
\mathbf{E}' = \mathbf{E}\, \cos \alpha + \mathbf{H}\, \sin \alpha\,, \qquad \mathbf{D}' = \mathbf{D}\, \cos \alpha + \mathbf{B}\,\sin \alpha\,
\end{equation} 
and
\begin{equation}\label{A12}
\mathbf{H}' = -\mathbf{E}\, \sin \alpha + \mathbf{H}\, \cos \alpha\,, \qquad \mathbf{B}' = -\mathbf{D}\, \sin \alpha + \mathbf{B}\, \cos \alpha\,,
\end{equation} 
constitutive relations~\eqref{A3} and Maxwell's equations in the absence of external sources remain invariant; in this connection, let us observe that  $\mathbf{F}'^{\,\pm} = \mathbf{F}^{\pm} \,e^{\mp\,i\,\alpha}$ and $\mathbf{S}'^{\,\pm} = \mathbf{S}^{\pm} \,e^{\mp\,i\,\alpha}$ under the duality rotations.

\subsection{Derivation of the Constitutive Relations}

For the sake of completeness, we briefly indicate here the derivation of constitutive relations~\eqref{A3}, which follow from the electric and magnetic parts of 
\begin{equation}\label{A13}
\sqrt{-g}\,F_{\mu \nu} = g_{\mu \alpha}\,g_{\nu \beta}\, \sqrt{-g}\,F^{\alpha \beta}\,,  \qquad   \sqrt{-g}\,F^{\mu \nu} = \sqrt{-g}\,g^{\mu \alpha}\,g^{\nu \beta}\, F_{\alpha \beta}\,,      
\end{equation} 
respectively. The electric part of the first relation in Eq.~\eqref{A13} implies
\begin{equation}\label{A14}
- \sqrt{-g}\, E_j = g_{00}\,\gamma_{jk}\,D_k - g_{jp}\,g_{0q}\, \epsilon^{pqn}\, H_n\,.           
\end{equation}
We multiply both sides of this equation by $g^{ij}/g_{00}$ and use $g^{ij}\,\gamma_{jk} = \delta^i_k$ as well as $g^{ij}\,g_{jp} = \delta^i_p - g^{0i}\,g_{0p}$ to obtain the first constitutive relation. 

Next, the magnetic part of the second relation in Eq.~\eqref{A13} can be written as
\begin{equation}\label{A15}
H_j  =  \sqrt{-g}\,\epsilon_{jmn}\,[g^{mp}\,g^{0n}\,E_p+ \frac{1}{2}\,\epsilon_{abc}\,g^{ma}\,g^{nb}\,B_c]\,.          
\end{equation}
Starting with $(g^{\mu \nu})$, one can determine the inverse matrix element $g_{0k}$ via
\begin{equation}\label{A16}
g \, \epsilon_{abc}\,g^{ai}\,g^{bj}\,g^{0c} = - \epsilon^{ijk}\,g_{0k}\,.
\end{equation}      
Furthermore, from the definition of the determinant and Eq.~\eqref{A7} we find 
\begin{equation}\label{A17}
\epsilon_{abc}\,g^{ai}\,g^{bj}\,g^{ck} = \epsilon^{ijk}\, \det(g^{pq}) = \epsilon^{ijk}\,\frac{g_{00}}{g}\,.
\end{equation}  
Multiplying both sides of Eq.~\eqref{A15} by $- \sqrt{-g}\, g^{ij}/g_{00}$ and using Eqs.~\eqref{A16} and~\eqref{A17}, we get the second constitutive relation.

\section*{Acknowledgments}

Discussions initiated by Friedrich W. Hehl  about parity violation in nonlocal gravity motivated the present investigation. I am also grateful to him for his insightful and critical comments.

\end{document}